\documentclass[11pt]{article}
\usepackage[dvips]{graphicx}
\usepackage{psfrag}
\usepackage{amssymb}
\usepackage{amsmath}
\usepackage{amscd}
\usepackage{latexsym}

\catcode`@=11
\renewcommand{\section}{\@startsection{section}{1}{0pt}{\medskipamount}
{\medskipamount}{\large\bf}}
\numberwithin{equation}{section}
\newcounter{saveeqn}

\catcode`@=12

\newcommand{\RR}{{\mathbb{R}}}
\newcommand{\CC}{{\mathbb{C}}}
\newcommand{\pa}{\partial}
\newcommand{\ii}{{\rm i}}
\newcommand{\dd}{{\rm d}}
\newcommand{\tp}{{\!\top}}
\newcommand{\sfrac}[2]{{\textstyle\frac{#1}{#2}}}

\newcommand{\SU}{\mathrm{SU}}
\newcommand{\Sp}{\mathrm{Sp}}
\newcommand{\UU}{\mathrm{U}}
\newcommand{\GL}{\mathrm{GL}}
\newcommand{\F}{\mathcal{F}}
\newcommand{\A}{\mathcal{A}}

\renewcommand{\=}{\ =\ }
\newcommand{\und}{\qquad\textrm{and}\qquad}

\begin{document}

\begin{titlepage}
\setcounter{page}{0}
\begin{flushright}
ITP--UH--13/09\\
\end{flushright}

\vskip 2.0cm

\begin{center}

{\Large\bf
Yang-Mills flows on nearly K\"ahler manifolds and $G_2$-instantons
}

\vspace{12mm}

{\large Derek~Harland${}^\dagger$,\ Tatiana~A.~Ivanova${}^*$,\
Olaf~Lechtenfeld${}^\dagger$,\ and\ Alexander~D.~Popov${}^*$
}
\\[8mm]
\noindent ${}^*${\em Bogoliubov Laboratory of Theoretical Physics, JINR\\
141980 Dubna, Moscow Region, Russia}\\
{Email: ita, popov@theor.jinr.ru}
\\[8mm]
\noindent ${}^\dagger${\em Institut f\"ur Theoretische Physik,
Leibniz Universit\"at Hannover \\
Appelstra\ss{}e 2, 30167 Hannover, Germany }\\
{Email: harland, lechtenf@itp.uni-hannover.de}

\vspace{12mm}

\begin{abstract}
\noindent
We consider Lie($G$)-valued $G$-invariant connections on bundles
over spaces $G/H$, $\RR{\times}G/H$ and $\RR^2{\times}G/H$, where
$G/H$ is a compact nearly K\"ahler six-dimensional homogeneous space,
and the manifolds $\RR{\times}G/H$ and $\RR^2{\times}G/H$ carry $G_2$-
and Spin(7)-structures, respectively. By making a $G$-invariant ansatz, 
Yang-Mills theory with torsion on $\RR{\times}G/H$ is reduced to 
Newtonian mechanics of a particle moving in a plane with a quartic potential.
For particular values of the torsion, we find explicit particle trajectories,
which obey first-order gradient or hamiltonian flow equations. In two cases,
these solutions correspond to anti-self-dual instantons associated with 
one of two $G_2$-structures on $\RR{\times}G/H$. It is shown that both $G_2$-instanton 
equations can be obtained from a single Spin(7)-instanton equation 
on $\RR^2{\times}G/H$.
\end{abstract}

\end{center}
\end{titlepage}

\section{Introduction and summary}
\label{sec1}
\noindent
The Yang-Mills equations in two, three and four dimensions have been 
intensively studied both in physics and mathematics. In mathematics, this study
(i.e.\ projectively flat unitary connections and stable bundles in 
$d{=}2$~\cite{AB}, the Chern-Simons model and knot theory in $d{=}3$, 
instantons and Donaldson invariants~\cite{DK} in $d{=}4$) 
has yielded a lot of new results in differential and algebraic geometry. 
In particular, a crucial role in $d{=}4$ gauge theory is played by the 
first-order anti-self-duality equations, which on manifolds $\RR{\times}X^3$ 
are precisely the Chern-Simons gradient flow equations.
The program of extending familiar constructions in gauge theory, 
associated to problems in low-dimensional topology, to higher dimensions, 
was proposed in~\cite{DT} and developed 
in~\cite{Lewis, Thomas, Tian, Br, SaEarp, Hay, DS}.\footnote{
For more literature see references therein.} 
An important role in this investigation
is played by first-order gauge equations which are a generalisation of the
anti-self-duality equations in $d{=}4$ to higher-dimensional manifolds with
special holonomy (or, more generally, with $G$-structure~\cite{Salamon, Joy}). 
Such equations in $d{>}4$ dimensions were first introduced 
in~\cite{Corrigan:1982th} and further considered e.g.\ 
in~\cite{Ward84, DUY, CSC, Bau, Tian, DS, Popov}.
Some of their solutions were found e.g.\ 
in~\cite{group1, group2, Ivanova:2009yi}.

In physics, interest in Yang-Mills theories in dimensions greater than 
four grew essentially after the discovery of superstring theory, 
which contains supersymmetric Yang-Mills in the low-energy limit 
in the presence of D-branes as well as in the heterotic case. 
In particular, heterotic strings yield $d{=}10$ heterotic supergravity, 
which contains the ${\cal N}{=}1$ supersymmetric
Yang-Mills model as a subsector~\cite{GSW}. Supersymmetry-preserving
compactifications on spacetimes $M_{10{-}d}\times X^d$ 
with further reduction to $M_{10{-}d}$ 
impose the above-mentioned first-order BPS-type gauge equations
on $X^d$~\cite{Corrigan:1982th, GSW}.
Initial choices for the internal manifold $X^6$ were K\"ahler
coset spaces and Calabi-Yau manifolds, as well as manifolds with exceptional
holonomy group $G_2$ for $d{=}7$ and Spin(7) for $d{=}8$.  However, 
it was realised that Calabi-Yau compactifications suffer from the presence 
of many massless moduli fields in the resulting four-dimensional effective 
theories.\footnote{
K\"ahler cosets also lead to non-realistic effective theories.} 
This problem can be cured (at least partially) by allowing for non-trivial 
$p$-form fluxes on~$X^d$. String vacua
with $p$-form fields along the extra dimensions (`flux compactifications')
have been intensively studied in recent years (see e.g.~\cite{group3} for
reviews, and also the references therein).

Compactifications in the presence of fluxes can be described
in the language of $G$-structures on $d$-dimensional manifolds $X^d$:
SU(3)-structure for dimension $d{=}6$, $G_2$-structure for $d{=}7$ and 
Spin(7)-structure for $d{=}8$. In the definition of all these $G$-structures 
there enters a $(d{-}4)$-form $\Psi$ on~$X^d$.
Thus, we deal with internal manifolds of special geometry and consider 
the three-form field~$\mathcal{H}=\ast\dd\Psi$ as torsion, where $\ast$ 
denotes the Hodge star operator. In particular, in six dimensions these 
manifolds may be non-K\"ahler and sometimes even non-complex. 

Flux compactifications have been investigated primarily for type II strings
and to a lesser extent in the heterotic theories, despite their long history
\cite{hetold}. The number of torsionful geometries that can serve as a
background for heterotic string compactifications seems rather limited.
Among them there are six-dimensional nilmanifolds, solvmanifolds,
nearly K\"ahler and nearly Calabi-Yau coset spaces. The last two kinds
of manifolds carry a natural almost complex structure which is not
integrable (for a discussion of their geometry see
e.g.~\cite{Gray,WG,Xu, But, group4} and references therein).

In heterotic string compactifications one has the freedom to choose a gauge
bundle since the simple embedding of the spin connection into the gauge
connection is ruled out for compactifications with $\dd\mathcal{H}{\neq}0$.
For the torsionful backgrounds, the allowed gauge bundle is restricted by
the Bianchi identity for the torsion field (anomaly cancellation) and by the 
Donaldson-Uhlenbeck-Yau equations~\cite{DUY} for $d{=}6$ or the $G_2$-instanton
equations~\cite{DT} for $d{=}7$. The construction of such vector bundles over
$G_2$-manifolds of topology $\RR\times X^6$ is the subject of the present paper.

The only known examples of compact nearly K\"ahler six-manifolds are the four
coset spaces \ $\SU(3)/\UU(1){\times}\UU(1)$, \ $\Sp(2)/\Sp(1){\times}\UU(1)$, \
$G_2/\SU(3)=S^6$ \ and \ $\SU(2)^3/\SU(2)=S^3{\times}S^3$. On all four cosets
$G/H$ we have a torsion $\mathcal{H}=\ast\dd\omega$ for an almost K\"ahler
form $\omega$. We describe some solutions of the Donaldson-Uhlenbeck-Yau
equations for the gauge group~$G$ on these cosets. 
Our ansatz for a $G$-invariant connection is
parameterised by a complex number~$\phi$, and the solutions show
the 3-symmetry characteristic of all nearly K\"ahler spaces.

Next, we step up to seven dimensions, extending $G/H$ by
a real line~$\RR_\tau$, so that $\phi\to\phi(\tau)\in\CC$ in
our $G$-invariant ansatz. For the torsion \ $\mathcal{H}=
-\sfrac13\kappa_1\ast\!(\dd\tau\wedge\dd\omega)+\sfrac13\kappa_2\,\dd\omega$ \
with $\kappa_1,\kappa_2\in\RR$, our ansatz reduces the Yang-Mills equations to
Newton's equations $\ddot\phi=f(\phi)$ for a particle in the complex 
$\phi$~plane, subject to a 3-symmetric cubic force~$f$.
For $\kappa_2{=}0$, there exists a potential of $\phi^4$ type, so
$f\sim\sfrac{\pa V}{\pa\bar\phi}$, and an action can be formulated,
which surprisingly agrees with the torsionful Yang-Mills action on our ansatz.
Yet, even for $\kappa_2{\neq}0$, we construct an explicit solution.

In special instances, $\ddot\phi\sim\sfrac{\pa V}{\pa\bar\phi}$ is implied by
a flow equation $\dot\phi\sim\sfrac{\pa W}{\pa\bar\phi}$. This flow is gradient
or hamiltonian, depending on whether the proportionality is real or imaginary.
Among the complex $\phi(\tau)$ trajectories, finite-action kinks occur when
$(\kappa_1,\kappa_2)=(\pm 3,0)$ and $(-1,0)$, as solutions to the
gradient and hamiltonian flow, respectively. The corresponding connections are finite-action solutions of the
Yang-Mills equations on $\RR \times G/H$. By a duality transformation, which
relates solutions for different values of~$\kappa_1$, infinite-action
solutions to the Yang-Mills equations are presented as well.

The cases $(\kappa_1,\kappa_2)=(3,0)$ and $(-1,0)$ mentioned above have a
clear geometrical meaning. The corresponding gradient and hamiltonian flow equations for 
$\phi$ follow from seven-dimensional anti-self-duality conditions based on one of 
two $G_2$-structures, called the $G_2$-instanton equations.  Now, these both 
descend from anti-self-duality equations
based on the Spin(7)-structure of the eight-dimensional space
$\RR_\tau\times\RR_\sigma\times G/H$. For the gradient case one reduces
over~$\RR_\sigma$, while the hamiltonian case arises upon reduction
over~$\RR_\tau$. The $G_2$-instanton equations can themselves be interpreted as gradient 
and hamiltonian flows for a certain action functional on the space of all 
connections.  We do not know of a similar geometrical
interpretation for any other special value of the torsion.

\bigskip

\section{Nearly K\"ahler coset spaces}
\label{sec2}

\subsection{Basic definitions}
An SU(3)-structure on a six-manifold is by definition a reduction of the structure
group of the tangent bundle to SU(3).  Manifolds of dimension six with SU(3)-structure admit a set of canonical objects fixed by SU(3), consisting of an
almost complex structure $J$, a Riemannian metric $g$, a real two-form $\omega$ and
a complex three-form $\Omega$.  With respect to $J$, the forms $\omega$ and $\Omega$
are of type (1,1) and (3,0), respectively, and there is a compatibility condition,
$g(J\cdot,\cdot)=\omega(\cdot,\cdot)$.  With respect to the volume form $V_g$ of $g$, $\omega$
and $\Omega$ are normalised so that
\begin{equation}
\label{normalisation}
\omega\wedge\omega\wedge\omega \= 6V_g \und
\Omega\wedge\bar{\Omega} \= -8\ii V_g\ .
\end{equation}
A nearly K\"ahler six-manifold is an SU(3)-structure manifold such that
\begin{equation} 
\dd\omega \= 3\rho\,\mathrm{Im}\Omega \und 
\dd\Omega \= 2\rho\,\omega\wedge\omega 
\end{equation}
for some real non-zero constant $\rho$, proportional to the square of the scalar curvature (if $\rho$ was zero, the manifold would be
Calabi-Yau).  Nearly K\"ahler manifolds were first studied by Gray \cite{Gray}, and they solve the Einstein equations with positive cosmological constant.  More generally, six-manifolds with SU(3)-structure are classified by
their intrinsic torsion, and nearly K\"ahler manifolds form one particular
intrinsic torsion class.

There are only four known examples of compact nearly K\"ahler six-manifolds, 
and they are all coset spaces:
\begin{equation}
\label{coset spaces}
\begin{array}{cc}
\SU(3)/\UU(1){\times}\UU(1)\ , & \Sp(2)/\Sp(1){\times}\UU(1)\ ,\\
G_2/\SU(3)=S^6\ , & \SU(2)^3/\SU(2)=S^3\times S^3\ .
\end{array}
\end{equation}
Here $\Sp(1){\times} \UU(1)$ is chosen to be a non-maximal subgroup of $\Sp(2)$:
if elements of $\Sp(2)$ are written as $2\times2$ quaternionic matrices, then
elements of $\Sp(1){\times} \UU(1)$ are written $\mathrm{diag}(p,q)$, with
$p\in \Sp(1)$ and $q\in \UU(1)$.
Also, $\SU(2)$ is the diagonal subgroup of $\SU(2){\times} \SU(2){\times} \SU(2)$.
These coset spaces $G/H$ were named 3-symmetric by Wolf and Gray, because the subgroup $H$ is the
fixed point set of an automorphism $s$ of $G$ satisfying $s^3=\mathrm{Id}$
\cite{WG,But}.

The 3-symmetry actually plays a fundamental role in defining the canonical
structures on the coset spaces.  The automorphism $s$ induces an automorphism
$S$ of the Lie algebra $\mathfrak{g}$ of $G$, that is
$S:\mathfrak{g}\rightarrow\mathfrak{g}$ is linear and satisfies
\begin{equation}
\label{3 symmetry}
[SX,SY] \= S[X,Y] \qquad \forall X,Y\in\mathfrak{g}\ .
\end{equation}
The cosets under consideration are all reductive, which means that there is a
decomposition $\mathfrak{g}= \mathfrak{h} \oplus \mathfrak{m}$, where
$\mathfrak{h}$ is the Lie algebra of $H$ and $\mathfrak{m}$ satisfies
$[\mathfrak{h},\mathfrak{m}]\subset\mathfrak{m}$.  Actually, on $\SU(2)^3/\SU(2)$
there is a choice of subspaces $\mathfrak{m}$; we choose $\mathfrak{m}$ so that
it is orthogonal to $\mathfrak{h}$ with respect to the Cartan-Killing form in
this case.  The map $S$ acts trivially on $\mathfrak{h}$ and non-trivially on
$\mathfrak{m}$; one can define a map $J:\mathfrak{m}\rightarrow\mathfrak{m}$ by
\begin{equation}
S|_\mathfrak{m} \= -\sfrac{1}{2} + \sfrac{\sqrt{3}}{2} J \=
\exp\left( \sfrac{2\pi}{3} J \right)\ .
\end{equation}
The map $J$ satisfies $J^2=-1$ and provides the almost complex structure on $G/H$.

A natural quadratic form on $\mathfrak{m}$ is given by the Cartan-Killing form
of $\mathfrak{g}$,
\begin{equation}
\left\langle X,Y \right\rangle_\mathfrak{g} \=
- \mathrm{Tr}_\mathfrak{g} ({\rm ad}(X)\circ {\rm ad}(Y))\ .
\end{equation}
This extends to a $G$-invariant metric $g$ on $G/H$.
The (1,1)-form $\omega$ is fixed by its compatibility with $g$ and $J$,
and $\Omega$ is the unique suitably normalised $G$-invariant (3,0)-form.

\subsection{Lie algebra identities}
In calculations, it is useful to choose a basis $\{I_A\}$ for the Lie algebra
$\mathfrak{g}$.  We do so in such a way that $I_a$ for $a=1,\dots,6$ form a
basis for $\mathfrak{m}$ and $I_i$ for $i=7,\dots,\mathrm{dim}(G)$ yield a basis
for $\mathfrak{h}$.  The structure constants $f_{AB}^C$ are defined by
\begin{equation}
[I_A,I_B]\=f_{AB}^CI_C \qquad\textrm{with}\qquad 
f_{AC}^Df_{DB}^C\=\delta_{AB}\ ,
\end{equation}
where we have chosen the basis so that it is orthonormal
with respect to the Cartan-Killing form.
Then $f_{ABC}:=f_{AB}^D\delta_{DC}$ is totally antisymmetric.

The reductive property of the coset means that the structure constants $f_{aij}$
vanish.  The components $J_{ab}$ of the almost complex structure $J$ are defined
via $J(I_a)=J_{ab}I_b$.  Then the 3-symmetry property (\ref{3 symmetry}) implies
useful identities involving $J$: notably, the tensor
\begin{equation}
\label{f identity 1}
\tilde{f}_{abc}\ :=\ f_{abd}J_{dc}
\end{equation}
is totally antisymmetric; also
\begin{equation}
\label{f identity 2}
J_{cd}f_{adi} \= J_{ad}f_{cdi}\ .
\end{equation}

Another useful identity is
\begin{equation}
\label{f identity 3}
J_{ab} f_{abi} \= 0\ .
\end{equation}
We do not have a general proof of this identity, but we have verified it on each of
the four coset spaces.  It has the following interpretation: the action of $H$ on
$\mathfrak{m}$ defines an embedding of $H$ in $\GL(6,\RR)$.  It is easy to show that
$H$ fixes the quadratic form $\left\langle\cdot,\cdot \right\rangle_\mathfrak{g}$
and almost complex structure $J$; hence $H$ is contained in $\UU(3)\subset \GL(6,\RR)$.
The above identity merely asserts that $H\subset \SU(3)$.  Geometrically, this means
that the natural $H$-structure on $G/H$ is contained within the $\SU(3)$-structure.

Apart from $\left\langle\cdot,\cdot \right\rangle_\mathfrak{g}$, there are two
other natural quadratic forms on $\mathfrak{m}$:
\begin{eqnarray}
\left\langle X,Y \right\rangle_\mathfrak{m} &:=& -\mathrm{Tr}_\mathfrak{m}
(P_\mathfrak{m}\circ {\rm ad}(X)\circ P_\mathfrak{m}\circ {\rm ad}(Y))\ , \\
\left\langle X,Y \right\rangle_\mathfrak{h} &:=& - \mathrm{Tr}_\mathfrak{h}\,
(P_\mathfrak{h}\,\circ {\rm ad}(X)\circ P_\mathfrak{m}\circ {\rm ad}(Y))\ ,
\end{eqnarray}
where $P_\mathfrak{m}$ and $P_\mathfrak{h}$ denote the projections onto 
$\mathfrak{m}$ and $\mathfrak{h}$, respectively.  It is easy to show that
\begin{equation}
\left\langle \cdot,\cdot \right\rangle_\mathfrak{g} \=
\left\langle \cdot,\cdot \right\rangle_\mathfrak{m} +
2\left\langle \cdot,\cdot \right\rangle_\mathfrak{h}\ .
\end{equation}
Furthermore, on the coset spaces in question, one also has
\begin{equation}
\left\langle \cdot,\cdot \right\rangle_\mathfrak{m} \=
\sfrac{1}{3} \left\langle \cdot,\cdot \right\rangle_\mathfrak{g}\ .
\end{equation}
Hence, in terms of the structure constants,
\begin{equation}
\label{f identity 4}
f_{aci}f_{bci} \= f_{acd}f_{bcd} \= \sfrac{1}{3}\delta_{ab}\ .
\end{equation}
The proof of this identity will be deferred until the end of this section.  Note that for three of the four coset spaces this identity has been verified directly in \cite{Lust:1986ix}.

\subsection{Orthonormal frame for the coset}
The metric and almost complex structure on $\mathfrak{m}$ lift to a $G$-invariant
metric and almost complex structure on $G/H$.  Local expressions for these can be
obtained by introducing an orthonormal frame as follows.  The basis elements $I_A$
of the Lie algebra $\mathfrak{g}$ can be represented by left-invariant vector
fields $\hat E_A$ on the Lie group $G$, and the dual basis $\hat{e}^A$ is a set
of left-invariant one-forms.  The space $G/H$ consists of left cosets $gH$ and 
the natural projection $g\mapsto gH$ is denoted $\pi:G\rightarrow G/H$.
Over a contractible open subset $U$ of $G/H$, one can choose a map $L:U\rightarrow G$
such that $\pi\circ L$ is the identity (in other words, $L$ is a local section of
the principal bundle $G\rightarrow G/H$).  The pull-backs of $\hat{e}^A$ by $L$
are denoted $e^A$.  In particular, $e^a$ form an orthonormal frame for $T^* (G/H)$
over $U$ (where again $a=1,\dots 6$), and we can write $e^i=e^i_ae^a$ with real
functions $e^i_a$.  The dual frame for $T(G/H)$ will be denoted $E_a$.
The forms $e^A$ obey the Maurer-Cartan equations,
\begin{equation}
\begin{aligned}
\label{MC}
\dd e^a &\=\ -f_{ib}^a\;e^i\wedge e^b\ -\ \sfrac12 f_{bc}^a\;e^b\wedge e^c\ , \\
\dd e^i &\=\!-\sfrac12 f_{bc}^i\,e^b\wedge e^c\ -\ \sfrac12 f_{jk}^i\,e^j\wedge e^k\ .
\end{aligned}
\end{equation}
Since all the connections we consider will be invariant under some action of $G$,
it will suffice to do calculations just over the subset $U$.

Local expressions for the $G$-invariant metric, almost complex structure, and
nearly K\"ahler form on $G/H$ are then
\begin{eqnarray}
g \= \delta_{ab}e^a e^b\ , \qquad
J \= J_{ab}e^a E_b\ , \und
\omega \= \sfrac{1}{2}J_{ab} e^a\wedge e^b.
\end{eqnarray}
One can also obtain a local expression for (3,0)-form $\Omega$. From 
(\ref{MC}) one can compute $\dd\omega$ and hence $\ast\dd\omega$:
\begin{equation}
\dd\omega \= -\sfrac{1}{2} \tilde{f}_{abc}\, e^a\wedge e^b\wedge e^c \und
\ast \dd\omega \= \sfrac{1}{2} f_{abc}\, e^a\wedge e^b\wedge e^c\ .
\end{equation}
We have that $\dd\omega= 3\rho\,\mathrm{Im}\Omega$, and $\Omega$ should be normalised
so that $\|\mathrm{Im}\Omega\|^2=4$.  On the other hand, from (\ref{f identity 4})
we compute that $\|\dd\omega\|^2 = 3$.  So it must be that $\rho = 1/2\sqrt{3}$ and
\begin{equation}
\label{Omega}
\mathrm{Im}\Omega \= 
-\sfrac{1}{\sqrt{3}} \tilde{f}_{abc}\,e^a\wedge e^b\wedge e^c\ , \qquad
\mathrm{Re}\Omega \= 
- \sfrac{1}{\sqrt{3}} f_{abc}\, e^a\wedge e^b\wedge e^c\ .
\end{equation}

Given a pair of differential forms $u,v$ such that the degree of $u$ is less than or equal to
the degree of $v$, their contraction is defined to be
\begin{equation}
u\lrcorner v\ :=\ \ast(u\wedge\ast v)\ .
\end{equation}
If $u$ and $v$ have the same degree, $u\lrcorner v$ coincides with the usual inner product of forms induced by the metric.  We are now in a position to prove (\ref{f identity 4}): from (\ref{Omega}), it is equivalent to
\begin{equation}
g( u\lrcorner\, \mathrm{Re}\Omega , v\lrcorner\, \mathrm{Re}\Omega) \= 2g(u,v) \quad \forall u,v\in \Lambda^1.
\end{equation}
This identity holds on any 6-manifold with SU(3)-structure, as can be verified by direct calculation in an orthonormal basis.

\bigskip

\section{Instantons in six dimensions}
\label{sec3}

\subsection{$\omega$-anti-self-duality}
Let $\Psi$ be a $(d{-}4)$-form on a $d$-dimensional Riemannian manifold.
A natural generalisation of the $d{=}4$ anti-self-duality equations
is the so-called $\Psi$-anti-self-duality equation,
\begin{equation}
\label{Omega SD}
\Psi\wedge \F \= -\ast \F\ .
\end{equation}
If $\Psi$ is closed, this equation implies the Yang-Mills equation, 
$D\ast \F=0$.
Equations of this sort were first written down in \cite{Corrigan:1982th}, using the
language of tensors rather than differential forms.  They often have an interpretation as BPS equations, in particular all of the $\Psi$-anti-self-duality equations considered in this paper are BPS equations.

On a nearly K\"ahler six-manifold, a natural choice for $\Psi$ is the (1,1)-form
$\omega$, giving
\begin{equation}
\label{omega SD}
\omega\wedge \F \= -\ast \F \qquad\Leftrightarrow\qquad
\ast(\omega\wedge \F) \= -\F\ .
\end{equation}
Of course, $\omega$ is not closed, so (\ref{omega SD}) does not imply the Yang-Mills
equation, but rather the Yang-Mills equation with torsion,
\begin{equation}
\label{YM torsion 1}
D\ast \F + \dd\omega\wedge \F \= 0\ .
\end{equation}
The $\omega$-anti-self-duality equation~(\ref{omega SD}) means that we are 
looking for eigen-two-forms $\F$ of the operator $\ast(\omega\wedge \cdot)$,
with eigenvalue $\lambda=-1$. The space $\Lambda^2$ of two-forms decomposes 
into three eigenspaces $\Lambda^2_\lambda$, with the following properties:
\begin{equation}
\begin{tabular}{|c|ccc|}
\hline
$\lambda$ & $2$ & $1$ & $-1$ \\
dim $\Lambda^2_\lambda$ & $1$ & $6$ & $8$ \\
$\F$-type & $\quad\sim\omega\quad$ & $(2,0)$, $(0,2)$ & $(1,1)\perp\omega$\\
\hline
\end{tabular}
\end{equation}
Hence, (\ref{omega SD}) is equivalent to the so-called Donaldson-Uhlenbeck-Yau,
or Hermitian-Yang-Mills, equations~\cite{DUY}:
\begin{equation}
\label{DUY equation}
\F^{0,2}\=\F^{2,0}\=0 \und \omega\lrcorner\,\F \= 0\ .
\end{equation}

It is interesting to note that when $\F$ solves the $\omega$-anti-self-duality
equation (\ref{omega SD}), the torsional term in the Yang-Mills equation
(\ref{YM torsion 1}) vanishes, as was pointed out by Xu \cite{Feng Xu}.
This is because $\F$ is a (1,1)-form and $\dd\omega$ is a sum of (3,0)- and
(0,3)-forms: their wedge product then has to vanish.

\subsection{Gauge group $H$}
First we consider $H$-instantons on $G/H$.  The natural projection $G\rightarrow G/H$
defines a principal bundle with structure group $H$, on which $G$ acts from the left.
There is a unique $G$-invariant connection on this bundle, the so-called canonical
connection \cite{Kobayashi-Nomizu1,MuellerHoissen:1987}.  On $S^2=\SU(2)/\UU(1)$
the canonical connection is the Dirac monopole and on $S^4=\Sp(2)/\Sp(1){\times} \Sp(1)$
it is the sum of an instanton and an anti-instanton, so it seems a good candidate
solution to (\ref{omega SD}) on a nearly K\"ahler coset space.

In local coordinates, the canonical connection is written
\begin{equation} 
\A \= e^i\,I_i \= e^a\,e_a^i\,I_i\ . 
\end{equation}
Its curvature $\F=\dd\A+\A\wedge \A$ is given in \cite{Kobayashi-Nomizu1}, chapter II,
theorem 11.1, and is also easily computed using (\ref{MC}):
\begin{equation} 
\F \= -\sfrac{1}{2} f_{ab}^i\,e^a\wedge e^b\,I_i\ . 
\end{equation}
The identity (\ref{f identity 2}) implies that this $\F$ is a (1,1)-form, since
(1,1)-forms $\theta$ are defined by the property
$\theta(J\cdot,J\cdot)=\theta(\cdot,\cdot)$.  The identity (\ref{f identity 3})
tells us that $\omega\lrcorner\,\F=0$, where $\omega=\frac{1}{2}J_{ab}e^a\wedge e^b$.
So on each of the four nearly K\"ahler coset spaces, the canonical connection
satisfies the Donaldson-Uhlenbeck-Yau equation (\ref{DUY equation}), or equivalently
the $\omega$-anti-self-duality equation (\ref{omega SD}).  The case $G/H=G_2/\SU(3)$ was
considered by Xu \cite{Feng Xu}, who also showed that the canonical connection admits
no continuous deformation preserving (\ref{omega SD}).  In other words, this connected
component of the moduli space of solutions consists of just a point.

\subsection{Gauge group $G$}
Next, we consider $G$-instantons on $G/H$.  According to \cite{Kobayashi-Nomizu1}
and \cite{Zoupanos}, $G$-invariant connections with gauge group $G$ are determined
by linear maps $\Lambda:\mathfrak{m}\rightarrow\mathfrak{g}$ which commute with the
adjoint action of $H$: $\Lambda(\mathrm{Ad}(h)X)=\mathrm{Ad}(h)\Lambda(X),\
\forall h\in H,\ X\in\mathfrak{m}$.  Such a linear map is represented by a matrix
$(\Phi_{aB})$, such that $\Lambda(I_a) = \Phi_{aB}I_B$, and in local coordinates the
connection is written
\begin{equation}\label{ansatz}
\A \= e^i\,I_i\ +\ e^a\,\Phi_{aB}I_B\ .
\end{equation}
We make the simple choice
\begin{equation}\label{Phichoice}
\Phi_{ab} \= \phi_1\,\delta_{ab}\ +\ \phi_2\,J_{ab} \und \Phi_{ai} \= 0 \ ,
\end{equation}
for real numbers $\phi_1$ and $\phi_2$, which is more general than the choice considered
in \cite{Ivanova:2009yi}.  On the space $G_2/\SU(3)$, (\ref{ansatz}) with
(\ref{Phichoice}) is the most general $G_2$-invariant connection, 
but on the other coset spaces it is not -- we will briefly discuss more general choices in the next section.
The curvature $\F=\frac{1}{2}\F_{ab}\,e^a\wedge e^b$ is given in
\cite{Kobayashi-Nomizu1}, chapter II, theorem 11.7, and can also be computed
using (\ref{MC}):
\begin{equation}
\label{curvature}
\F_{ab} \= f_{ac}^i (\Phi^\tp\Phi-\mathrm{Id})_{cb}\,I_i\ +\
f_{abc} ( -\Phi+(\Phi^\tp)^2)_{cd}\,I_d\ ,
\end{equation}
where $(\mathrm{Id},J)^\tp=(\mathrm{Id},-J)$.
Note that $\F$ remembers the 3-symmetry $S$:
\begin{equation}
\Phi\ \mapsto\ \exp(\sfrac23\pi J)\Phi \qquad\Rightarrow\qquad
(-\Phi+(\Phi^\tp)^2)\ \mapsto\ \exp(\sfrac23\pi J)(-\Phi+(\Phi^\tp)^2)\ .
\end{equation}

The two-forms 
\begin{equation}
e^c\lrcorner\ast\dd\omega\=\sfrac32f_{abc}\,e^a\wedge e^b \und
e^c\lrcorner\,\dd\omega\= -\sfrac32\tilde{f}_{abc}\,e^a\wedge e^b
\end{equation}
are clearly of
type (2,0)+(0,2), since $\dd\omega$ is of type (3,0)+(0,3).  So this connection
solves the $\omega$-anti-self-duality equation (\ref{omega SD}) if and only if 
\begin{equation}
-\Phi+(\Phi^\tp)^2\=0\ .
\end{equation}
Apart from the canonical connection $\Phi=0$,
the other solutions to this equation are
\begin{equation}
\Phi \= \mathrm{Id}\ ,\quad \exp(\sfrac23\pi J)\ ,\quad \exp(\sfrac43\pi J)\ .
\end{equation}
Note that these connections in fact all have zero curvature.

\bigskip

\section{Yang-Mills equations in seven dimensions}
\label{sec4}

\subsection{From Yang-Mills theory to a $\Phi^4$ model}
On a $d$-dimensional Riemannian manifold, the Yang-Mills equation with torsion is
\begin{equation}
\label{YM torsion 2}
D *\F\ +\ *\mathcal{H}\wedge \F \= 0\ ,
\end{equation}
with $\mathcal{H}$ a three-form.  Equation (\ref{YM torsion 1}) is a special case in $d=6$.
We will study solutions of this equation on the seven-dimensional manifolds
$\RR \times G/H$, with $G/H$ a nearly K\"ahler coset space.  We choose the metric
and volume form,
\begin{equation}
g_7 \= (e^0)^2 + g_6 \und V_7 \= e^0\wedge V_6\ ,
\end{equation}
where $e^0=\dd\tau$ and $\tau$ is a coordinate on $\RR$, while
$g_6$ and $V_6$ are the metric and volume form on $G/H$.  
For $\mathcal{H}$ we make the choice
\begin{equation}
*\mathcal{H} \= -\sfrac13\kappa_1\,\dd\tau\wedge\dd\omega\ +\
\sfrac13\kappa_2\,*\dd\omega\ .
\end{equation}
This choice for $\mathcal{H}$ is clearly invariant under the action of $G$, and under translations in and reversals of $\tau$ -- in fact, it is the most general possible choice satisfying these conditions.  If one does not require invariance under $\tau$-reversals, then a term proportional $\dd\tau\wedge\omega$ could be added to $\mathcal{H}$ and possibly others, depending on the choice of coset space.

For the connection one-form $\A=\A_0e^0+\A_ae^a$, we copy from the
previous section the $G$-invariant ansatz (\ref{ansatz}) and (\ref{Phichoice}),
\begin{equation} \label{ansatz2}
\A(\tau) \= e^i\,I_i\ +\ e^a\,\Phi_{ab}(\tau)\,I_b \qquad\textrm{with}\qquad
\Phi \= \phi_1\,\mathrm{Id}\ +\ \phi_2\,J\ ,
\end{equation}
where $\phi_1$ and $\phi_2$ are now functions of $\tau$.
This ansatz has $\A_0=0$, but no generality is lost here 
since such a gauge can always be chosen. The curvature 
$\F=\F_{0a}\,e^0\wedge e^a+\sfrac12\F_{ab}\,e^a\wedge e^b$
of this connection has the components (see~(\ref{curvature}))
\begin{equation}
\F_{ab} \= f_{ac}^i (\Phi^\tp\Phi-\mathrm{Id})_{cb}\,I_i\ +\
f_{abc} ( -\Phi+(\Phi^\tp)^2)_{cd}\,I_d \und
\F_{0a} \= \dot{\Phi}_{ab}\,I_b\ ,
\end{equation}
where a dot denotes a derivative with respect to $\tau$.

In order to write the Yang-Mills equation in components, it is necessary to introduce
the torsionful spin connection on $G/H$ \cite{Zoupanos}.  Recall that a linear
connection is a matrix of one-forms $\omega^a_b=e^c\omega^a_{cb}$.  The connection is
metric compatible if $\omega_a^c g_{cb}$ is anti-symmetric, and its torsion is a vector
of two-forms $T^a=\frac12T^a_{bc}\,e^b\wedge e^c$ determined by the structure equation
\begin{equation}
\label{Maurer-Cartan}
\dd e^a\ +\ \omega^a_b \wedge e^b \= T^a\ .
\end{equation}
Our choice is
\begin{equation}
T^a \= - e^a\lrcorner\, \mathcal{H} \qquad \Leftrightarrow \qquad
T^a_{bc} \= \kappa_{ad}\,f_{dbc} \qquad\textrm{with}\qquad
\kappa\ :=\ \kappa_1\,\mathrm{Id}\ -\ \kappa_2\,J\ .
\end{equation}
We take $\omega^a_b$ to be the
unique metric-compatible linear connection with this torsion.  Explicitly,
\begin{equation}
\omega^a_{cb}\=e^i_cf^a_{ib}\ +\ \sfrac12(\kappa+\mathrm{Id})_{ad} f_{dcb}\ .
\end{equation}
The torsionful spin connection on $\RR\times G/H$ is given by a similar formula,
with additional components vanishing:
\begin{equation}
\omega^0_{0b} \= \omega^a_{0b} \= \omega^0_{cb} \= 0\ .
\end{equation}

Using (\ref{Maurer-Cartan}), one can show that the Yang-Mills equation with torsion
(\ref{YM torsion 2}) is equivalent to
\begin{eqnarray}
\label{YM components 1}
E_a \F^{a0}\ +\ \omega^a_{ab} \F^{b0}\ +\ [\A_a,\F^{a0}] &=& 0 \ ,\\
\label{YM components 2}
E_0\F^{0b}\ +\ E_a\F^{ab}\ +\ \omega^d_{da}\F^{ab}\ +\ \omega^b_{cd}\F^{cd}\ +\
[ \A_a, \F^{ab}] &=& 0\ .
\end{eqnarray}
It is now a matter of computation to substitute the ansatz (\ref{ansatz2}) into
(\ref{YM components 1}) and (\ref{YM components 2}), making use of structure constant
identities introduced above.  One finds that (\ref{YM components 1}) is identically
satisfied, while (\ref{YM components 2}) is equivalent to
\begin{equation}
\label{Phi equation}
6\,\ddot{\Phi} \= (\kappa{-}1)\,\Phi\ -\ (\kappa{+}3)(\Phi^\tp)^2\ +\
4\,\Phi^\tp\Phi^2\ .
\end{equation}

For more general choices of connection, equation (\ref{YM components 1}) is not automatically solved.  For example, in the cases $G/H={\rm SU}(3)/{\rm U}(1){\times} {\rm U}(1)$ and $\SU(2)^3/\SU(2)$, the most general $G$-invariant connection is parametrised by three complex scalars $\psi_1,\psi_2,\psi_3$.\footnote{For $G/H{=}\Sp(2)/\Sp(1){\times}\UU(1)$ one has $\psi_1=\psi_2$; for $S^6$ and one has $\psi_1=\psi_2=\psi_3=\phi$.}  Equation (\ref{YM components 1}) then reads
\begin{equation}
\dot\psi_1\bar\psi_1 - \dot{\bar{\psi}}_1\psi_1 \=
\dot\psi_2\bar\psi_2 - \dot{\bar{\psi}}_2\psi_2 \=
\dot\psi_3\bar\psi_3 - \dot{\bar{\psi}}_3\psi_3.
\end{equation}
The general solution to these equations is difficult to find, but one obvious solution is $\psi_1=\psi_2=\psi_3$.  This returns us to our original ansatz (\ref{ansatz2}).  More general ans\"atze will be discussed in a future publication.

\subsection{Action}
There is an alternative method of deriving (\ref{Phi equation}), which involves
working with actions rather than equations of motion.  
Notice that, with $\kappa_2=0$,
the Yang-Mills equation with torsion (\ref{YM torsion 2}) 
is the equation of motion for the action
\begin{equation}
\label{action torsion}
S \= \int_{\mathbb{R}\times G/H} \mathrm{Tr} \Bigl[ \F\wedge* \F\ +\
\sfrac{1}{3}\kappa_1\,\dd\tau\wedge \omega \wedge \F \wedge \F \Bigr]\ .
\end{equation}
In the Calabi-Yau ($\rho=0$) limit this action agrees with the standard 
Yang-Mills action up to a boundary term.  
Substituting the ansatz (\ref{ansatz2}) into this action gives
\begin{eqnarray}
S &=& \textrm{Vol}(G/H) \int_{\mathbb{R}} \!\dd\tau\ \mathrm{Tr} \left[
2\,\dot{\Phi}^\tp\dot{\Phi}\ +\ \widehat{V}(\Phi) \right] 
\qquad\textrm{with} \\[6pt]
3\,\widehat{V}(\Phi) &=& 
(1{-}\sfrac{\kappa_1}{3})\,\mathrm{Id}\ +\
(\kappa_1{-}1)\,\Phi^\tp\Phi\ -\ 
(1{+}\sfrac{\kappa_1}{3}) \bigl(\Phi^3+(\Phi^\tp)^3\bigr)\ +\ 
2\,(\Phi^\tp\Phi)^2 \ .\label{Phipot}
\end{eqnarray}
The Euler-Lagrange equation for this integral is again (\ref{Phi equation}).
The matrix-valued potential~$\widehat{V}$ is invariant under the action of the 
$S_3$ permutation group generated by the 3-symmetry and the conjugation~$\tp$.

Let us review what we have done.  Starting from the action (\ref{action torsion}),
we first substituted an ansatz and then derived an equation of motion.  Previously,
we derived the equation of motion (\ref{YM torsion 2}) for the action and then
substituted the ansatz.  There is no reason to expect these two procedures to lead
to the same differential equation, unless the ansatz chosen is the most general ansatz
invariant under a given symmetry (this is called the principle of symmetry
criticality \cite{Manton-Sutcliffe}).  In our case, the two procedures \emph{did}
lead to the same differential equation.  However, our ansatz, while $G$-invariant,
is certainly not the most general $G$-invariant ansatz, except in the case of $S^6$.  That the two procedures lead to the same differential equation on the other coset spaces could perhaps be attributed to the algebraic similarities between all four coset spaces.

\bigskip

\section{Solutions of the Yang-Mills equation}
\label{sec5}

\subsection{Critical points of the potential}
Throughout this section, we identify the matrix-valued function
$\Phi=\phi_1 \mathrm{Id} + \phi_2 J$ with the complex-valued function
$\phi=\phi_1 + \ii\phi_2$ and, likewise, 
interpret $\kappa$ as a complex number.  
Since $\widehat{V}(\Phi)^\tp=\widehat{V}(\Phi^\tp)=\widehat{V}(\Phi)$
in~(\ref{Phipot}), we also define
\begin{equation} \label{phipot}
\widehat{V}(\Phi)\ =:\ V(\phi)\,\mathrm{Id} \qquad\Rightarrow\qquad
3\,V(\phi)\= (1{-}\sfrac{\kappa_1}{3})\ +\ (\kappa_1{-}1)\,|\phi|^2\ -\ 
(1{+}\sfrac{\kappa_1}{3})\,2\,\mathrm{Re}\,\phi^3\ +\ 2\,|\phi|^4\ .
\end{equation}

If $\kappa_2=0$, the equation of motion
(\ref{Phi equation}) can be written in terms of the real function $V$,
\begin{equation} \label{phi equation}
6\,\ddot{\phi}\=
(\kappa{-}1)\,\phi\ -\ (\kappa{+}3){\bar\phi}^2\ +\ 4\,\bar\phi\phi^2\=
3\,\frac{\partial V}{\partial \bar{\phi}}\ .
\end{equation}
This is the equation of motion of a particle moving in the complex plane under 
the influence of a potential $-V$.  Equation (\ref{phi equation}) admits this
mechanical interpretation only when $\kappa_2=0$ since, 
if $\kappa$ was complex, the potential function $V$ could not be chosen real.  
In this section, we study solutions of (\ref{phi equation}) 
with $\kappa_2=0$, using this mechanical analogy.
We briefly discuss solutions of (\ref{Phi equation}) with $\kappa_2\neq0$ at the end of the section.
Figure~1 displays the equipotential lines of $V(\phi)$ for two special cases.
\begin{figure}[ht]
\centerline{
\includegraphics[width=7cm]{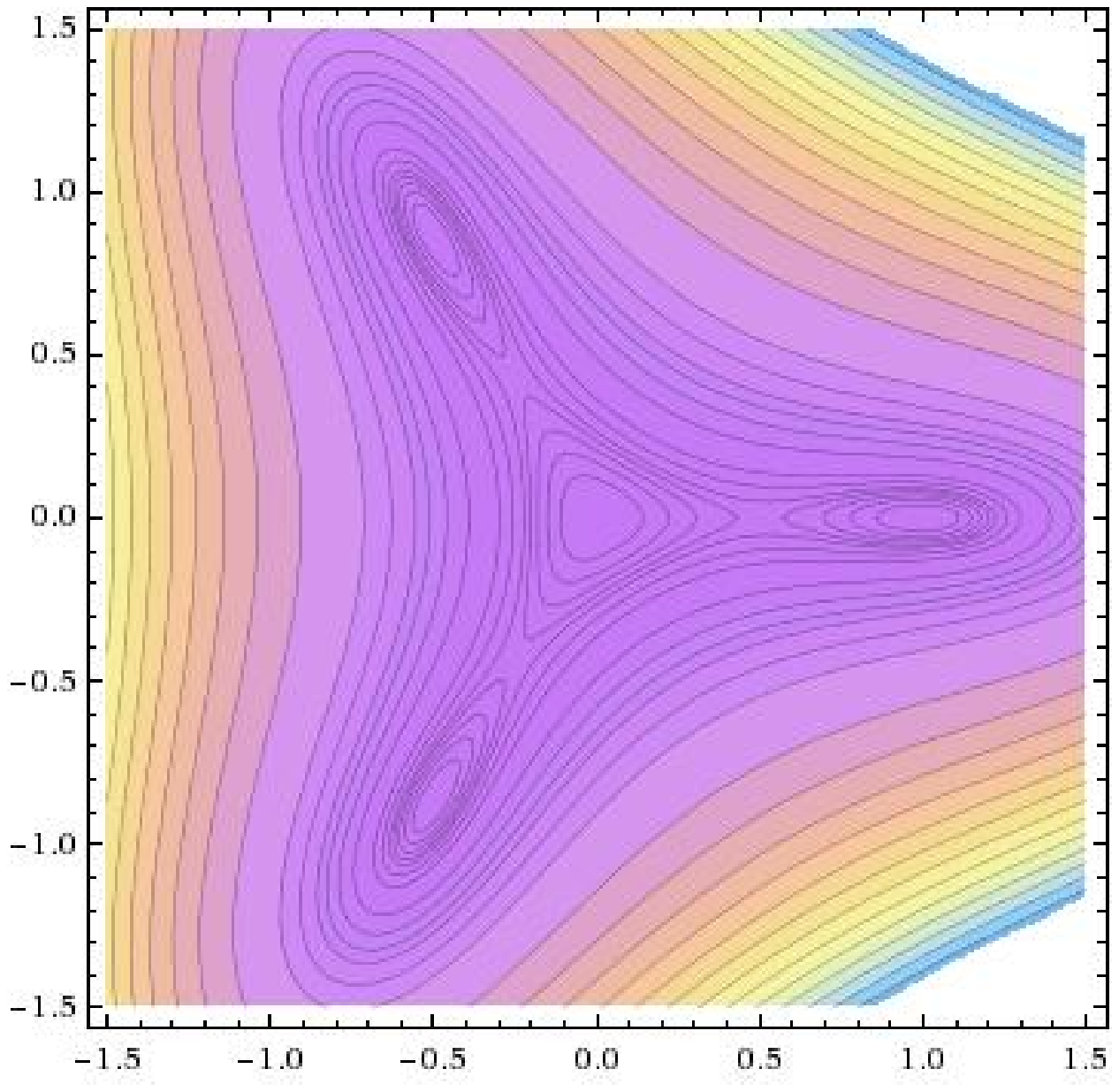}
\hfill
\includegraphics[width=7cm]{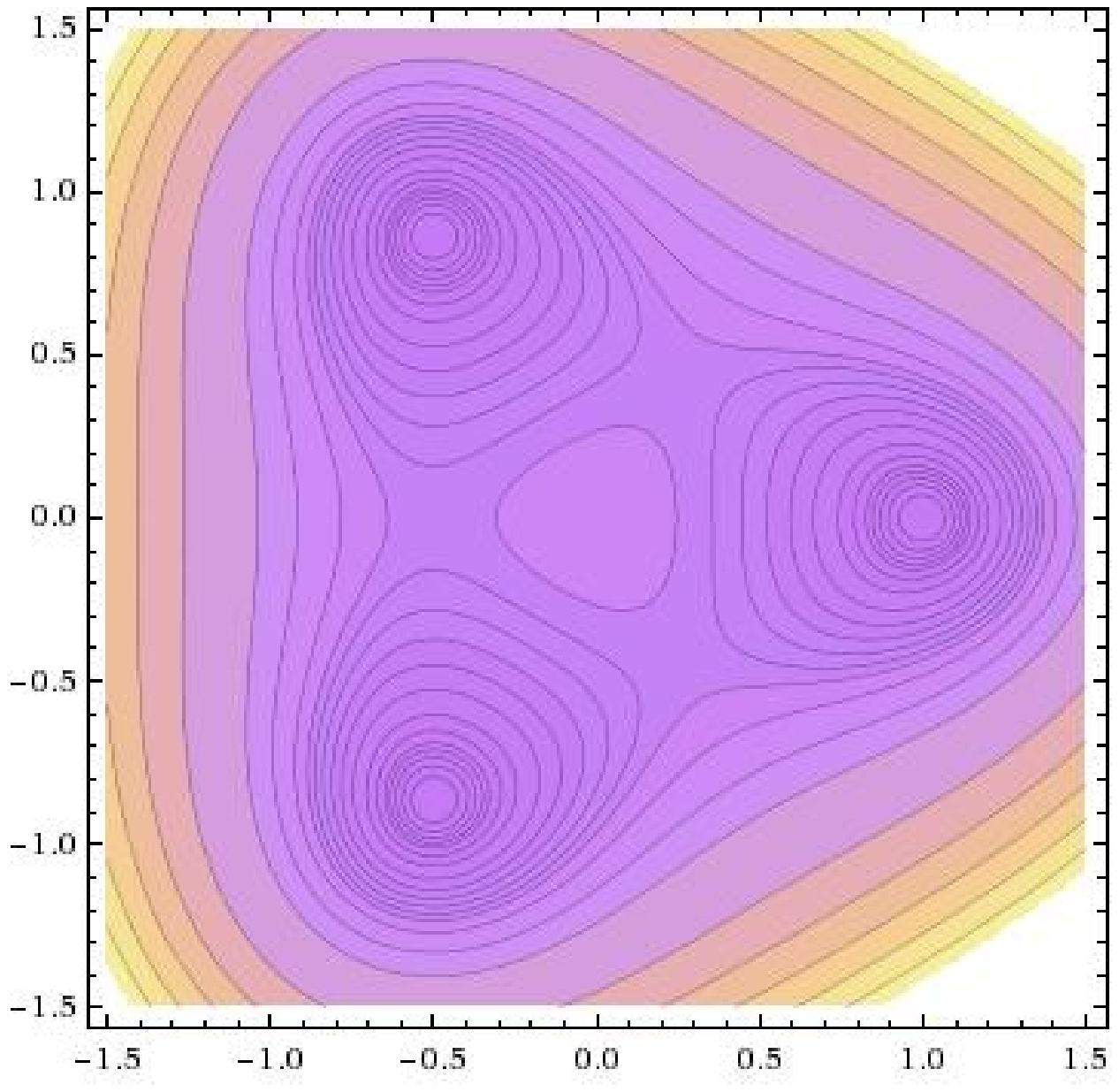}
}
\caption{contour plots of the potential $V(\phi)$ 
for $\kappa=+3$ (left) and for $\kappa=-1$ (right)}
\label{fig:1}
\end{figure}

We are particularly interested in instantons, which correspond to
particle trajectories interpolating between critical points of $V$, attained at
$\tau=\pm\infty$.  By conservation of energy, such a trajectory can exist only
if $V|_{+\infty}=V|_{-\infty}$.
The critical points $\phi^0$ of $V$ along the real axis are
\begin{equation}
\begin{tabular}{|c|ccc|}
\hline
$\phi^0$ & 0 & 1 & $\nu$ \\[4pt]
$V(\phi^0)$ & $\ \ \sfrac29(1{-}2\nu)\ \ $ & 0 & 
$\sfrac29(1{+}\nu)(1{-}\nu)^3$ \\[2pt]
\hline
\end{tabular}
\qquad \textrm{with} \quad \nu\ :=\ \sfrac14(\kappa_1{-}1)\ .
\end{equation}
Since $V$ is invariant under 
$\phi\mapsto\exp(\frac23\pi\ii)\phi$, for nonzero $\phi^0$ there are 
further critical points
$\exp(\frac23\pi\ii)\phi^0$ and $\exp(\frac43\pi\ii)\phi^0$, 
degenerate in energy with $\phi^0$.
At any value of $\kappa_1$, one may therefore search for trajectories 
connecting two critical points related by 3-symmetry,
which we shall call ``transverse''. We will show below that
transverse trajectories exist for $\kappa_1=-7,-1$ (i.e.~$\nu=-2,-\sfrac12$).
If $\kappa_1=-3,3,9$ (i.e.~$\nu=-1,\sfrac12,2$), 
two of the critical points on the real axis are degenerate in energy, 
and one may in addition look for ``radial'' trajectories connecting them.

The sought-for instanton configurations have finite action only when
$V|_{\pm\infty}=0$. Among the five special cases just mentioned, 
this occurs for $\kappa_1=-3,-1,3$ (i.e.~$\nu=-1,-\sfrac12,\sfrac12$).
Finite-energy bounce solutions, connecting $\phi^0=0$ to itself,
may exist for $\kappa_1<-3$ and for $3<\kappa_1<5$.
\begin{equation}
\begin{tabular}{|c|ccccc|}
\hline
$\kappa_1$    & $-7$      & $-3$   & $-1$        & $3$        & $9$     \\
$\nu$         & $-2$      & $-1$   & $-\sfrac12$ & $\sfrac12$ & $2$     \\
degeneration  & none & $V(\exp(\ii\alpha))$ & none & $V(0)=V(1)$ & $V(0)=V(2)$ \\
instanton     &transverse & radial & transverse  & radial     & radial  \\
$\phi^0(\pm\infty)$ & $\exp(\pm\frac23\pi\ii)(-2)$ & $\exp(\ii\alpha)(\pm1)$ & 
$\exp(\pm\frac23\pi\ii)(+1)$ & $\sfrac12\pm\sfrac12$ & $1\pm1$ \\
action        & infinite  & finite & finite      & finite     & infinite\\
\hline
\end{tabular}
\end{equation}

\subsection{Duality}
When $\kappa_2=0$, there is a surprising duality that relates pairs of values of
$\kappa_1$.  This is best seen when the equation of motion (\ref{phi equation})
and the potential~(\ref{phipot}) are rewritten in terms of $\nu$:
\begin{equation}
6\,\ddot{\phi} \= 4\nu\,\phi\ -\ 4(\nu{+}1)\,\bar{\phi}^2\ +\
4\,\bar{\phi}\phi^2 \und
3\,V(\phi)\= \sfrac23(1{-}2\nu)\ +\ 4\nu\,|\phi|^2\ -\ 
\sfrac83(1{+}\nu)\,\mathrm{Re}\,\phi^3\ +\ 2\,|\phi|^4\ .
\end{equation}
It is straightforward to check that 
\begin{equation}
\bigr(\,\nu\,,\,\phi(\tau)\,\bigl)\ \ \mapsto\ 
\bigl(\,\sfrac1\nu\,,\,\sfrac1\nu\,\phi(\sfrac\tau\nu)\,\bigr)
\end{equation}
maps solutions of (\ref{phi equation}) to other solutions.
We do not know the origin of this duality.

\subsection{Gradient flow}
We discuss here the case $(\kappa_1,\kappa_2)=(3,0)$. 
Our discussion applies
also to $(\kappa_1,\kappa_2)=(9,0)$, via the duality transformation.  
The potential $V$ can be written in terms of a real ``superpotential'' $W$:
\begin{equation} \label{W}
3\,V \= 2 \left| \frac{\partial W}{\partial \bar{\phi}} \right|^2 
\qquad\textrm{with}\qquad
W \= \sfrac{1}{3}(\phi^3{+}\bar{\phi}^3) - |\phi|^2\ .
\end{equation}
So (\ref{phi equation}) is implied by the gradient flow equation
\begin{equation}
\label{gradient flow}
\pm\sqrt{3}\,\dot{\phi} \= \bar{\phi}^2-\phi 
\= \frac{\partial W}{\partial \bar{\phi}}\ .
\end{equation}
Finite-action kink solutions are
\begin{equation}
\phi(\tau) \= 
\sfrac12 \bigl(1\pm\tanh(\sfrac{\tau{-}\tau_0}{2\sqrt{3}})\bigr)\ ,
\end{equation}
with $\tau_0$ being the collective coordinate.  
Further solutions are obtained by applying the
3-symmetry.  Since $W$ is 3-symmetric, it is clear that the gradient flow
(which reduces the value of $W$ along a path) does not have any transverse solutions.
\begin{figure}[ht]
\centerline{\includegraphics[width=10cm]{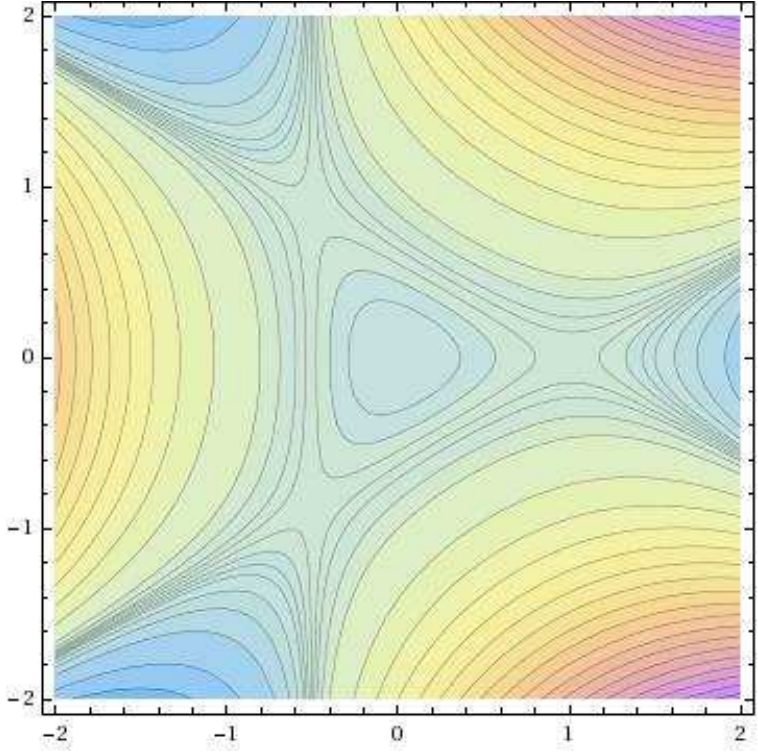}}
\caption{contour plot of the superpotential $W(\phi)$ 
for $\kappa=+3$ (and for $\kappa=-1$)}
\label{fig:2}
\end{figure}

We have also found explicit infinite-action solutions of the gradient flow equations:
if we define two real functions $r(\tau)$ and $\varphi(\tau)$ by the polar decomposition
$\phi=r\exp(\ii\varphi)$, then (\ref{gradient flow}) is equivalent to
\begin{equation}
\sqrt{3}\, \dot r      \= r - r^2 \cos 3\varphi \und
\sqrt{3}\, \dot\varphi \= r \sin 3\varphi\ .
\end{equation}
It follows that
\begin{equation}
\frac{\dd r}{\dd\varphi} \= \mathrm{cosec}\, 3\varphi - r \cot 3\varphi \ ,
\end{equation}
assuming $\dot{\varphi} \neq0$.  This can be integrated using a standard formula
to give
\begin{equation}
r(\varphi) \= -\sfrac{1}{3}\,(\sin 3\varphi)^{-1/3} \left[ \cos 3\varphi \ \, 
{}_2 F_1\!\left(\sfrac12,\sfrac56,\sfrac32,\cos^2 3\varphi\right)\ +\ C \right]
\end{equation}
for some real integration constant $C$.  
The hypergeometric function ${}_2F_1$ arises
from the antiderivative of $(\sin 3\varphi)^{-2/3}$.
Since $r(\varphi)$ diverges for $3\varphi=n\pi$, the trajectories are unbounded.
However, this solution does not capture the special case of radial motion:
\begin{equation}
\dot\varphi = 0 \quad \Leftrightarrow \quad 3\varphi=n\pi\quad\textrm{and}\quad
\sqrt{3}\,\dot r = r(1{-}r)\ ,
\end{equation}
which yields our previous kinks, moving radially in the special directions.

\subsection{Continuous symmetry}
The case $(\kappa_1,\kappa_2)=(-3,0)$ is special: 
firstly, because it is fixed by the duality
transformation; and secondly, because the potential function $V$ is invariant
under not only the 3-symmetry, but also under U(1) rotations of $\phi$.  Again, we can
find a superpotential $W$,
\begin{equation} \label{superpotential -3}
3\,V \= 2 \left| \frac{\partial W}{\partial \bar{\phi}} \right|^2 
\qquad\textrm{with}\qquad
W \= \sfrac{2}{3}|\phi|^3-2|\phi|\ .
\end{equation}
Like before, solutions of the gradient flow equation 
\begin{equation}
\pm\sqrt{3}\,\dot{\phi} \= \sfrac{\phi}{|\phi|}\,\bigl(1-|\phi|^2\bigr) 
\= \frac{\partial W}{\partial \bar{\phi}}
\end{equation}
solve (\ref{phi equation}). 
However, care should be taken near the origin, where $W$ is not differentiable.
The finite-action solutions are
\begin{equation}
\phi(\tau) \= \pm\tanh(\sfrac{\tau{-}\tau_0}{\sqrt{3}})
\end{equation}
and U(1) rotations of these.  The only other solutions are infinite-action
trajectories connecting the critical circle $|\phi|=1$ with $\phi=\infty$:
\begin{equation}
\phi(\tau) \= \mp\coth(\sfrac{\tau{-}\tau_0}{\sqrt{3}})\ ,
\end{equation}
modulo U(1) rotations.
In particular, there are no transverse solutions of the gradient flow equation.

\subsection{Hamiltonian flow}
Now we consider transverse trajectories. Without loss of generality, we look for
transverse trajectories connecting $\exp(-\sfrac23\pi\ii)\phi^0$ and $\exp(\sfrac23\pi\ii)\phi^0$
with $\phi^0$ a non-zero real critical point, i.e.~$\phi^0=1$ or $\phi^0=\nu$.  For simplicity, we assume that $\phi_1$
is constant along these trajectories.  Then $\phi_1$ and $\kappa$ should be chosen so
that $\ddot\phi_1=0$ for all $\phi_2$. Since
\begin{equation}
\label{dV/dphi1}
6\,\ddot\phi_1 \=
\left[ (\kappa_1{-}1)\phi_1-(\kappa_1{+}3)\phi_1^2+4\phi_1^3 \right]\ +\
\kappa_2(2\phi_1{+}1)\,\phi_2\ +\
\left[ (\kappa_1{+}3)+4\phi_1 \right]\,\phi_2^2\ ,
\end{equation}
we get exactly three solutions for $\kappa_2=0$:
\begin{equation} \label{3cases}
(\phi_1,\kappa_1)\=(0,-3)\,,\ (-\sfrac12,-1)\,,\ (1,-7)\ .
\end{equation}
We will return to $\kappa_2\neq0$ solutions momentarily.

The case $\kappa_1=-3$ was treated above and yields radial trajectories.
The cases $\kappa_1=-1$ and $\kappa_1=-7$ are related by the duality 
transformation, so we consider here just $(\kappa_1,\kappa_2)=(-1,0)$. 
Then (\ref{phi equation}) is implied by a first-order hamiltonian flow,
\begin{equation}
\label{Hamiltonian flow}
\pm\sqrt{3}\,\dot\phi\=\ii(\bar{\phi}^2-\phi)
\=\ii\,\frac{\partial W}{\partial\bar{\phi}}
\qquad\textrm{with}\qquad
W \= \sfrac{1}{3}(\phi^3{+}\bar{\phi}^3) - |\phi|^2\ ,
\end{equation}
i.e.\ the hamiltonian is exactly the superpotential of (\ref{W}).  
There are finite-action solutions
\begin{equation}
\phi(\tau)\=
-\sfrac12\ \pm\ \ii\sfrac{\sqrt{3}}{2}\tanh(\sfrac{\tau{-}\tau_0}{2})\ .
\end{equation}
Two further solutions are obtained on application of the 3-symmetry.

It is straightforward to write down infinite-action solutions of
(\ref{Hamiltonian flow}), at least in implicit form.  The value of $W$ is conserved
by the flow, so solutions $\phi(\tau)$ of the flow obey $W(\phi(\tau))=C$ for
some constant~$C$.  In polar coordinates $\phi=r\exp(\ii\varphi)$, this reads
\begin{equation}
\cos3\varphi \= \frac{3}{2}\,\frac{r^2+C}{r^3}\ ,
\end{equation}
which yields $r(\varphi)$ as a solution of a cubic equation.
In particular, for $-1/3<C<0$ there are periodic trajectories.
It is amusing to note that the gradient-flow and hamiltonian-flow cases
are related by flipping the sign of~$\nu$.

\subsection{Solutions without action}
Referring again to equation (\ref{dV/dphi1}) we see that, 
if $\kappa_2\neq0$, then $\ddot\phi_1=0$ enforces $\phi_1=-\sfrac12$,
hence the case
\begin{equation}
(\phi_1,\kappa_1)\=(-\sfrac12,-1)
\end{equation}
of (\ref{3cases}) allows us to turn on~$\kappa_2$.
With these values fixed, (\ref{Phi equation}) is equivalent to
\begin{equation}
6\,\ddot\phi_2 \= 4\,(\phi_2-\sfrac{\sqrt{3}}{2})\,
(\phi_2+\sfrac{\sqrt{3}}{2})\,(\phi_2-\sfrac{\kappa_2}{4})\ .
\end{equation}
This equation has kink-type solutions whenever the roots of the polynomial on
the right hand side are evenly spaced. 
This occurs not only in the case $\kappa_2=0$ discussed above, 
but also when $\kappa_2=\pm6\sqrt{3}$.  The corresponding kink solutions are
\begin{equation}
\phi \= -\sfrac12\ +\ 
\epsilon\,\ii\bigl(\sfrac{\sqrt{3}}{2}\pm\sqrt{3}\tanh(\tau{-}\tau_0)\bigr)
\qquad\textrm{with}\quad \epsilon=\textrm{sgn}(\kappa_2)\ .
\end{equation}

\bigskip

\section{Instanton equations in seven and eight dimensions}
\label{sec6}

\subsection{Anti-self-duality in eight dimensions}
In the previous section we constructed solutions of the Yang-Mills equation on
$G/H\times\RR$ for special values of $\kappa$.  We found that the second-order
Yang-Mills equations actually reduced to first-order equations for these special
values.  In this section we will show that those first-order equations which
admit finite-energy instantons have a natural geometrical interpretation: 
they take the anti-self-duality form (\ref{Omega SD}) with a suitably chosen 
three-form $\Psi$.

It is most convenient to start in eight dimensions rather than in seven.  
Let $x^7$ and $x^8$ denote coordinates on $\RR^2$ and let $e^7=\dd x^7$ and
$e^8=\dd x^8$; then the forms
\begin{equation}
\widetilde{\omega} \= \omega\ +\ e^7\wedge e^8 \und
\widetilde{\Omega} \= \Omega\wedge(e^7+\ii e^8)
\end{equation}
define an SU(4)-structure on $G/H\times\RR^2$.  
The associated metric and volume form are
\begin{equation}
g_8 \= g_6 + (e^7)^2 + (e^8)^2 \und V_8\=V_6\wedge e^7\wedge e^8\ .
\end{equation}
The four-form
\begin{equation}
\Sigma \= \sfrac{1}{2}\,\widetilde{\omega}\wedge\widetilde{\omega}\ -\
\mathrm{Re}\widetilde{\Omega}
\end{equation}
defines a Spin(7)-structure. The operator $*_8(\Sigma\wedge\cdot)$ on two-forms
has eigenvalues -1 and 3, with eigenspaces of dimensions 21 and 7, respectively.
So it makes sense to consider the $\Sigma$-anti-self-duality equation
\begin{equation}
\label{Sigma SD}
\Sigma\wedge \F \= -*_8 \F.
\end{equation}

This equation has been studied in \cite{Corrigan:1982th,Lewis,DT,DS}.
With respect to a complex basis
\begin{equation}
\{\Theta^\alpha\}: \quad\
\Theta^1 = e^1+\ii e^2\ ,\quad
\Theta^2 = e^3+\ii e^4\ ,\quad
\Theta^3 = e^5+\ii e^6\ ,\quad
\Theta^4 = e^7+\ii e^8
\end{equation}
it reads
\begin{equation} \label{complex SD}
\widetilde\omega\ \lrcorner\ \F \= 0 \und
\F_{\bar{\alpha}\bar{\beta}} \= -\sfrac{1}{2}
\epsilon_{\bar{\alpha}\bar{\beta}\bar{\gamma}\bar{\delta}}
\F^{\bar{\gamma}\bar{\delta}} \quad \textrm{(6 real equations)}\ ,
\end{equation}
where we have raised indices using the almost Hermitian metric: $\F^{\bar{\alpha}\bar{\beta}} =
\F_{\alpha\beta}\,g^{\alpha\bar\alpha}g^{\beta\bar\beta}$ with $g^{\alpha\bar\alpha}=\delta^{\alpha\bar\alpha}$.
For $\A_7=\A_8=0$ the equations (\ref{complex SD}) reduce to
\begin{equation}
(\partial_7 +\ii\,\partial_8)\,\A_{\bar{p}}\=
\epsilon_{\bar{p}\bar{q}\bar{r}}\,\F^{\bar{q}\bar{r}}\quad\mbox{for } p,q,r=1,2,3\ ,
\end{equation}
whose stable points satisfy $\F^{0,2}=0$.  In the real basis, 
(\ref{complex SD}) read
\begin{equation}
 \partial_7\A_a - J_{ab}\,\partial_8\A_b\=\sqrt{3} f_{abc}\F_{bc}\ .
\end{equation}

\subsection{Gradient flow}
Now we step down to seven dimensions.  Consider the seven-manifold
$G/H\times\RR$, with $\RR$ parametrised by $x^7$, the metric $g_7$ induced
from $g_8$ and the volume form $V_7=V_6\wedge e^7$.  
Then our four-form $\Sigma$ descends as follows, 
\begin{equation}
\Sigma \= \Xi \wedge e^8\ +\ *_7\Xi\ ,
\end{equation}
where
\begin{equation}
\Xi \= \omega\wedge e^7\ +\ \mathrm{Im}\Omega \und
*_7 \Xi \= \sfrac{1}{2}\,\omega\wedge\omega\ -\ \mathrm{Re}\Omega\wedge e^7
\end{equation}
live on $G/H\times\RR$.  The three-form $\Xi$ defines a $G_2$-structure, 
which is compatible with the metric in the sense that
$*_7(i_u\Xi\wedge i_v\Xi\wedge\Xi)=6\,g_7(u,v)$ for all tangent vectors
$u,v$~\cite{bryant:1987}.

Associated to $\Xi$ is an anti-self-duality equation (\ref{Omega SD}).
The eigenvalue problem for the operator $*_7(\Xi\wedge\cdot)$ on two-forms $\F$
is characterised as follows,
\begin{equation}
\begin{tabular}{|c|cc|} 
\hline
$\lambda$ & $2$ & $-1$ \\
dim $\Lambda^2_\lambda$ & $7$ & $14$ \\
$\F$-type & $\qquad i_u\Xi\qquad$ & $*_7\Xi\wedge \F=0$ \\
\hline
\end{tabular}
\end{equation}
The space $\Lambda^2_{-1}$ maps to the Lie algebra of $G_2$ under 
the isomorphism $\Lambda^2\cong \mathfrak{so}(7)$.

Now suppose that $\F$ is a connection on $G/H\times\RR^2$ pulled back from
$G/H\times \RR$ or, equivalently, that $\A_8=0$ and $\A_1,\dots,\A_7$ 
are independent of $x^8$ in some gauge.  Then it is easy to show that
\begin{equation}
\Sigma\wedge \F \= \Xi\wedge\F\wedge e^8\ +\ *_7\Xi\wedge\F \und
*_8\F \= *_7\F\wedge e^8\ .
\end{equation}
Hence, the $\Sigma$-anti-self-duality (\ref{Sigma SD}) in eight dimensions
descends to
\begin{equation}
\label{Xi SD}
\Xi\wedge\F \= -*_7\F\ .
\end{equation}
This equation was studied in detail in \cite{SaEarp}.  Differentiating, one sees that this equation implies the Yang-Mills equation 
(\ref{YM torsion 2}) with torsion given by $(\kappa_1,\kappa_2)=(3,0)$, on
identifying $\tau=x^7$.  Further, a solution of the anti-self-duality equation
(\ref{omega SD}) on $G/H$ pulls back to a $\tau$-independent solution
of (\ref{Xi SD}).

Let us rewrite (\ref{Xi SD}) in components.  This is easily done
using the fact that (\ref{Xi SD}) is equivalent to $(i_u\Xi)\lrcorner\,\F=0$ 
for tangent vectors $u=E_1,\dots,E_7$. With
\begin{equation}
i_{E_7}\Xi\=-\sfrac{1}{2} J_{ab} e^a\wedge e^b \und
i_{E_a}\Xi\=-J_{ab} e^b\wedge e^7\ +\ 
\sqrt{3}\tilde{f}_{abc}e^b\wedge e^c\ ,
\end{equation}
(\ref{Xi SD}) is equivalent to
\begin{equation}
\label{Xi SD comp}
J_{ab}\F_{ab} \= 0 \und
\F_{7a} \= \sqrt{3} f_{abc}\F_{bc}\ .
\end{equation}
The second relation is the flow equation introduced in \cite{Ivanova:2009yi}.
We have shown in Section~\ref{sec3} that the first relation is satisfied by our
ansatz (\ref{ansatz2}); substituting this ansatz into 
the second relation yields precisely the gradient flow equation 
(\ref{gradient flow}) for $\kappa=3$.

\subsection{Hamiltonian flow}
We now repeat the discussion of the previous subsection, but with the roles
of $x^7$ and $x^8$ reversed.  We regard $x^8$ as a coordinate on the factor $\RR$
of $G/H\times\RR$, denote by $g_7'$ the metric induced from $g_8$ and choose the
volume form $V_7'=V_6\wedge e^8$.  Then
\begin{equation}
\Sigma \= -\Xi' \wedge e^7\ +\ *_7'\Xi'\ ,
\end{equation}
where
\begin{equation}
\Xi' \= \omega\wedge e^8\ +\ \mathrm{Re}\Omega \und
*_7' \Xi' \= \sfrac{1}{2}\,\omega\wedge\omega\ +\ \mathrm{Im}\Omega\wedge e^8\ .
\end{equation}
Again, $\Xi'$ defines a $G_2$-structure on $G/H\times\RR$, and the action
of $*_7'(\Xi'\wedge\cdot)$ exactly mirrors that of $*_7(\Xi\wedge\cdot)$.  
In particular,
if $\F$ is a connection on $G/H\times\RR$ pulled back to $G/H\times\RR^2$,
then (\ref{Sigma SD}) is equivalent to
\begin{equation}
\label{Xi' SD}
\Xi'\wedge\F \= -*_7'\F\ .
\end{equation}
Which second-order equation is implied by (\ref{Xi' SD})?  
Differentiating, one obtains
\begin{equation}
D*_7'\F +(3\rho\,\mathrm{Im}\Omega\wedge e^8+2\rho\,\omega\wedge\omega)\wedge\F\=0\ .
\end{equation}
Taking into account the equivalence of (\ref{Xi' SD}) and $*_7'\Xi'\wedge\F=0$,
one arrives at the Yang-Mills equation~(\ref{YM torsion 2}) with torsion
for $(\kappa_1,\kappa_2)=(-1,0)$.

The component form of (\ref{Xi' SD}) is
\begin{equation}
\label{Xi' SD comp}
J_{ab}\F_{ab} \= 0 \und
\F_{8a} \= - \sqrt{3} \tilde{f}_{abc}\F_{bc}\ .
\end{equation}
With the ansatz (\ref{ansatz2}), the first equation is 
again automatically satisfied, while the second one
is exactly equivalent to the hamiltonian flow equation (\ref{Hamiltonian flow})
with $\kappa=-1$.

\subsection{Continuous symmetry}
The fixed points of the gradient flow or hamiltonian flow equations are
the critical points of the superpotential~$W$. For the special U(1)-symmetric
case $(\kappa_1,\kappa_2)=(-3,0)$, the superpotential (\ref{superpotential -3}) 
yields the fixed points $|\phi|^2=1$.
{}From the discussion in Section \ref{sec3} it is clear that these are 
the solutions of the $\omega$-self-duality equation,
\begin{equation}
*_6 \F \= \omega\wedge\F\ ,
\end{equation}
for a $G$-invariant connection on $G/H$. Indeed, differentiating this equation
gives the Yang-Mills equation (\ref{YM torsion 2}) with torsion via
$(\kappa_1,\kappa_2)=(-3,0)$.
This simple geometrical interpretation for the static solutions does not seem 
to extend to the general solutions of the flow equation~(\ref{gradient flow})
in this situation.  Note that the $\omega$-self-dual equation is not a BPS equation.

\subsection{First-order flows}

We close with a second interpretation of the $G_2$-instanton equations (\ref{Xi SD}) and (\ref{Xi' SD}), which accounts for the appearance of a superpotential with gradient and hamiltonian flows in section~\ref{sec5}.

In his thesis \cite{Feng Xu}, Xu noted that the $\omega$-anti-self-duality 
equation (\ref{omega SD}) is equivalent to
\begin{equation}
\label{vanishing torsion}
\dd\omega \wedge \F \=0\ .
\end{equation}
We have already seen that (\ref{omega SD}) implies (\ref{vanishing torsion}). 
To show the converse, we first observe that (\ref{vanishing torsion}) implies 
$\F^{0,2}=\F^{2,0}=0$.  It follows that
\begin{equation}
\mathrm{Re}\Omega \wedge\F \=0\ .
\end{equation}
Second, differentiating and applying the Bianchi identity yields
\begin{equation}
\omega\wedge\omega\wedge\F\=0 
\qquad\Leftrightarrow\qquad \omega\lrcorner\,\F\=0\ .
\end{equation}
Thus, $\F$ contains only a (1,1) part orthogonal to~$\omega$.
Since the Donaldson-Uhlenbeck-Yau equations~(\ref{DUY equation}) are equivalent
to~(\ref{omega SD}), this proves the assertion.

Xu then introduced an action
\begin{equation}
\label{superpotential}
\int_{X^6} \mathrm{Tr} ( \omega \wedge \F \wedge \F )
\end{equation}
for a connection on a nearly K\"ahler six-manifold $X^6$, 
whose equation of motion is (\ref{vanishing torsion}) 
and whose gradient flow equation is
\begin{equation}
\label{Xu flow}
\frac{\partial \A}{\partial\tau} \= *(\F\wedge\dd\omega)\ .
\end{equation}
Given a local orthonormal frame $e^a$ for the cotangent bundle of $X^6$,
we contract both sides with the $e^a$. Employing the identity
\begin{equation}
*(\F\wedge\dd\omega)\lrcorner e^a \= (*\dd\omega)\lrcorner(\F\wedge e^a)\ ,
\end{equation}
one sees that the flow equation is equivalent to
\begin{equation} \frac{\partial \A_a}{\partial \tau} \=
\sfrac{3}{2} f_{abc} \F_{bc}\ ,
\end{equation}
which coincides with the second equation of (\ref{Xi SD comp}) after a rescaling in $\tau$.

Thus, the $G_2$-instanton equation (\ref{Xi SD}) implies the gradient flow for the action (\ref{superpotential}).  This explains why (\ref{Xi SD}) reduces to a gradient flow equation for $\phi$.  Substituting the ansatz (\ref{ansatz2}) into (\ref{superpotential}) should give something proportional to the superpotential (\ref{W}), and this is easily verified:
\begin{equation}
\begin{aligned}
3 \int_{G/H} \mathrm{Tr} ( \omega \wedge \F \wedge \F ) &\= -\mathrm{Vol}(G/H)\, \mathrm{Tr} \left[ 1-3 \Phi^\tp\Phi + \Phi^3+(\Phi^\tp)^3 \right] \\
&\= -\mathrm{Vol}(G/H)\,(1+3W)\ .
\end{aligned}
\end{equation}
Similarly, the hamiltonian flow equation for (\ref{superpotential}) is
\begin{equation}
\label{Xu' flow}
\frac{\partial \A_a}{\partial\sigma}J_{ab}e^b \= *(\F\wedge\dd\omega)\ .
\end{equation}
This is equivalent to the second equation of (\ref{Xi' SD comp}), and reduces to the hamiltonian flow for $W$.

Finally, we comment on the relation between the gradient and hamiltonian flow equations and the remaining part of the $G_2$-instanton equations, $\omega\lrcorner\,\F=0$ (the first equation in (\ref{Xi SD comp}) or (\ref{Xi' SD comp})).  By employing the identities $J\ast \pa_\tau\A = \sfrac12 \pa_\tau\A\wedge\omega\wedge\omega$ and $J({\rm Im}\Omega\wedge \F)={\rm Re}\Omega\wedge \F$, the gradient flow (\ref{Xu flow}) can be rearranged to read
\begin{equation}
\frac{1}{2}\frac{\pa \A}{\pa \tau}\wedge\omega\wedge\omega = 3\rho\,{\rm Re}\Omega\wedge \F\ .
\end{equation}
By taking the exterior derivative, and using the fact that $D(\pa_\tau\A)=\pa_\tau\F$, this equation implies
\begin{equation}
\frac{\pa}{\pa \tau} (\omega\lrcorner\, \F) = 12\rho^2\, \omega\lrcorner\, \F\ .
\end{equation}
A similar argument shows that, for the hamiltonian flow (\ref{Xu' flow}), $\pa_\tau(\omega\lrcorner\,\F)=0$.  So, we should not be surprised that $\omega\lrcorner\,\F=0$ holds for our gradient and hamiltonian flows: if one requires this to hold at $\tau=\pm\infty$, it will hold everywhere.

It is a curious fact that a nearly K\"ahler structure can itself be regarded as a critical point of a hamiltonian flow \cite{Hitchin:2001rw} -- perhaps there is some connection between this and the gradient and hamiltonian flows described above.

\bigskip

\noindent
{\bf Acknowledgements}

\medskip

\noindent
We thank Christoph N\"olle for collaboration at an early stage.
This work was supported in part by the cluster of excellence EXC~201
``Quantum Engineering and Space-Time Research'',
by the Deutsche Forschungsgemeinschaft (DFG)
and by the Heisenberg-Landau program.
The work of T.A.I.\ and A.D.P.\ was partially supported by the
Russian Foundation for Basic Research (grant RFBR 09-02-91347).
The work of D.H.\ is supported by Graduiertenkolleg GRK~1463 
``Analysis, Geometry and String Theory''.

\end{document}